\documentclass[12pt,epsf]{article}
\usepackage{graphicx}
\usepackage{epsfig}
\begin{document}

\def\a{\alpha}
\def\b{\beta}
\def\c{\varepsilon}
\def\d{\delta}
\def\e{\epsilon}
\def\f{\phi}
\def\g{\gamma}
\def\h{\theta}
\def\k{\kappa}
\def\l{\lambda}
\def\m{\mu}
\def\n{\nu}
\def\p{\psi}
\def\q{\partial}
\def\r{\rho}
\def\s{\sigma}
\def\t{\tau}
\def\u{\upsilon}
\def\v{\varphi}
\def\w{\omega}
\def\x{\xi}
\def\y{\eta}
\def\z{\zeta}
\def\D{\Delta}
\def\G{\Gamma}
\def\H{\Theta}
\def\L{\Lambda}
\def\F{\Phi}
\def\P{\Psi}
\def\S{\Sigma}

\def\o{\over}
\def\beq{\begin{eqnarray}}
\def\eeq{\end{eqnarray}}
\newcommand{\gsim}{ \mathop{}_{\textstyle \sim}^{\textstyle >} }
\newcommand{\lsim}{ \mathop{}_{\textstyle \sim}^{\textstyle <} }
\newcommand{\vev}[1]{ \left\langle {#1} \right\rangle }
\newcommand{\bra}[1]{ \langle {#1} | }
\newcommand{\ket}[1]{ | {#1} \rangle }
\newcommand{\EV}{ {\rm eV} }
\newcommand{\KEV}{ {\rm keV} }
\newcommand{\MEV}{ {\rm MeV} }
\newcommand{\GEV}{ {\rm GeV} }
\newcommand{\TEV}{ {\rm TeV} }
\def\diag{\mathop{\rm diag}\nolimits}
\def\Spin{\mathop{\rm Spin}}
\def\SO{\mathop{\rm SO}}
\def\O{\mathop{\rm O}}
\def\SU{\mathop{\rm SU}}
\def\U{\mathop{\rm U}}
\def\Sp{\mathop{\rm Sp}}
\def\SL{\mathop{\rm SL}}
\def\tr{\mathop{\rm tr}}

\def\IJMP{Int.~J.~Mod.~Phys. }
\def\MPL{Mod.~Phys.~Lett. }
\def\NP{Nucl.~Phys. }
\def\PL{Phys.~Lett. }
\def\PR{Phys.~Rev. }
\def\PRL{Phys.~Rev.~Lett. }
\def\PTP{Prog.~Theor.~Phys. }
\def\ZP{Z.~Phys. }

\newcommand{\beqr}{\begin{array}}  
\newcommand {\eeqr}{\end{array}}
\newcommand{\la}{\left\langle}  
\newcommand{\ra}{\right\rangle}
\newcommand{\non}{\nonumber}  
\newcommand{\ds}{\displaystyle}
\newcommand{\red}{\textcolor{red}}
\def\ubl{U(1)$_{\rm B-L}$}
\def\REF#1{(\ref{#1})}
\def\lrf#1#2{ \left(\frac{#1}{#2}\right)}
\def\lrfp#1#2#3{ \left(\frac{#1}{#2} \right)^{#3}}
\def\OG#1{ {\cal O}(#1){\rm\,GeV}}

\def\J{{\cal J}}
\def\aa{{\dot \a}}
\def\bb{{\dot \b}}
\def\ss{{\bar \s}}
\def\hh{{\bar \h}}
\def\Z{{\cal Z}}


\baselineskip 0.7cm

\begin{titlepage}

\begin{flushright}
UT-12-01\\
IPMU-12-0014
\end{flushright}

\vskip 1.35cm
\begin{center}
{\large \bf
On the Trace Anomaly and the Anomaly Puzzle \\ in ${\cal N}=1$ Pure Yang-Mills}
\vskip 1.2cm
Kazuya Yonekura$^{1,2}$ 
\vskip 0.4cm

{\it $^1$ Institute for the Physics and Mathematics of 
the Universe (IPMU),\\ 
University of Tokyo, Chiba 277-8568, Japan\\
$^2$  Department of Physics, University of Tokyo,\\
    Tokyo 113-0033, Japan}

\vskip 1.5cm

\abstract{ 
The trace anomaly of the energy-momentum tensor is usually quoted in the form which is proportional to the beta function of the theory.
However, there are in general many definitions of gauge couplings depending on renormalization schemes, and hence many beta functions.
In particular, ${\cal N}=1$ supersymmetric pure Yang-Mills has
the holomorphic gauge coupling whose beta function is one-loop exact, and the canonical gauge coupling whose beta function is given by the 
Novikov-Shifman-Vainshtein-Zakharov beta function.
In this paper, we study which beta function should appear in the trace anomaly in ${\cal N}=1$ pure Yang-Mills. 
We calculate the trace anomaly by employing the ${\cal N}=4$ regularization of ${\cal N}=1$ pure Yang-Mills.
It is shown that the trace anomaly is given by one-loop exact form if the composite operator appearing in the trace anomaly
is renormalized in a preferred way. 
This result gives the simplest resolution to the anomaly puzzle in ${\cal N}=1$ pure Yang-Mills.
The most important point is to examine in which scheme the quantum action principle is valid, which is crucial in the derivation
of the trace anomaly.

}
\end{center}
\end{titlepage}

\setcounter{page}{2}

\section{Introduction}
\label{sec:1}

In supersymmetric field theories, any operator is contained in some supermultiplet. 
This is also the case for the energy-momentum tensor $T^{\m\n}$. It was discovered long time ago~\cite{Ferrara:1974pz} that
the energy-momentum tensor is contained in a supermultiplet called the supercurrent ${\cal J}^\m$. 
The components of the supercurrent is given as
\beq
{\cal J}^\m =j^\m_R +\h S^\m +{\rm h.c.}+2\h \s_\n \bar{\h} T^{\m\n}+\cdots,
\eeq
where $j_R^\m$ is an R-symmetry current and $S^\m$ is the current of supersymmetry.

In the past, it was believed~\cite{Clark:1978jx} that the conservation equation of the energy-momentum tensor is extended to 
the superspace equation given by
\beq
\bar{D}^{\dot{\a}} {\cal J}_{\a\dot{\a}} \approx D_\a X  \label{eq:chiral}
\eeq
where $X$ is some chiral superfield, or else 
\beq
\bar{D}^{\dot{\a}} {\cal J}_{\a\dot{\a}} \approx -\frac{1}{4}{\bar D}^2 D_\a J, \label{eq:linear}
\eeq
where $J$ is some real vector superfield, and we use the symbol $\approx$ for equations which are valid
if equations of motion are used (i.e., the equations are valid up to contact terms when they are inserted into correlation functions). 
Eq.~(\ref{eq:linear}) is possible only if a conserved R-symmetry current exists, so
we focus on Eq.~(\ref{eq:chiral}) in the following.

Eq.~(\ref{eq:chiral}) gives the conservation equations of $T^{\m\n}$ and $S^\m$, but it also gives the additional constraint given by
\beq
\frac{2}{3} T^\m_\m+i \q_\m j^\m_R \approx F_X, \label{eq:anomaly}
\eeq
where $F_X$ is the $F$-component of $X$. The fact that the right hand side of Eq.~(\ref{eq:anomaly}) is given by a single $F$-term
causes a problem called the anomaly puzzle~\cite{Grisaru:1978vx}. The axial anomaly contributing to $\q_\m j^\m_R$ is often said to be one-loop exact
due to the Adler-Bardeen theorem~\cite{Adler:1969er}. This one-loop exactness seems to indicate that the chiral superfield $X$ is one-loop exact.
On the other hand, the trace anomaly $T^\m_\m$ usually receives higher order corrections. Then, higher order corrections to $X$ seem to be
necessary.
These two statements look inconsistent with each other. This is the anomaly puzzle.
Many works have been done to solve the anomaly puzzle, 
including Refs.~\cite{Grisaru:1985yk,Grisaru:1985ik,Shifman:1986zi,ArkaniHamed:1997mj} and references therein.

The situation have been changed by the discovery of a new supercurrent multiplet~\cite{Komargodski:2010rb}. 
(See also Ref.~\cite{Magro:2001aj} for an early work.) It has been found that the supercurrent equation can be relaxed to the form
\beq
\bar{D}^{\dot{\a}} {\cal J}_{\a\dot{\a}} \approx D_\a X-\frac{1}{4}{\bar D}^2 D_\a J.
\eeq
Then, Eq.~(\ref{eq:anomaly}) is now modified to
\beq
\frac{2}{3} T^\m_\m & \approx &{\rm Re}(F_X)-\frac{1}{6}D_J, \\
\q_\m j^\m_R &\approx & {\rm Im}(F_X),
\eeq
where $D_J$ is the $D$-component of $J$.
These equations suggest that $X$ can be one-loop exact to maintain the one-loop exactness of the axial anomaly,
and higher order corrections to the trace anomaly is accounted for by the new term $D_J$. Based on this observation,
a resolution to the anomaly puzzle has been proposed~\cite{Yonekura:2010mc}.
(See also Ref.~\cite{Huang:2010tn} for a related work.)
More explicitly, let us consider a theory with matter chiral fields $\F_r$ which transform in the representation $r$ of the gauge group.
The gauge field strength chiral field is denoted as $W_\a$, and the superpotential is denoted as $W$. Then,
$X$ and $J$ are given as
\beq
 X&=& \frac{4}{3}\left[3W-\sum_r \F_r\frac{\q W}{\q \F_r} -\frac{3t_{Ad}-\sum_r t_r}{32\pi^2}W^\a W_\a \right], \label{eq:anompart} \\
 {J}&=&-4\sum_{r}\g_r \F_r^\dagger e^{-2V} \F_r , \label{eq:correctionpart}
\eeq
where $t_{Ad}$ and $t_r$ are the Dynkin indices for the adjoint representation and the representation $r$, respectively,
$\g_r$ is the matter anomalous dimension, and the summation is over all matter fields. 
Several consistency checks have been done for this proposal in Ref.~\cite{Yonekura:2010mc}.

However, there still remains a puzzle. The problem already exists in ${\cal N}=1$ supersymmetric pure Yang-Mills,
and we restrict our attention to this case. In this theory, there is no candidate for the operator $J$, and hence the 
situation regarding the anomaly puzzle of this theory have not been changed.
The trace anomaly is often quoted in the form
\beq
T^\m_\m \approx -\frac{\b}{4g^4}F_{\m\n}F^{\m\n} \label{eq:usualtraceanom}
\eeq
where $\b$ is the beta function of the gauge coupling, and $F_{\m\n}$ is the gauge field strength.
In supersymmetric theories, there is a known ``exact'' beta function~\cite{Novikov:1983uc,Shifman:1986zi,ArkaniHamed:1997mj}, called the
Novikov-Shifman-Vainshtein-Zakharov (NSVZ) beta function. In the case of the ${\cal N}=1$ pure Yang-Mills, it is given as
\beq
\b_{\rm NSVZ}=\m \frac{\q g^2}{\q \m}=-\frac{1}{8\pi^2}\frac{3t_{Ad}g^4}{1-t_{Ad}g^2/8\pi^2}.
\eeq
If this beta function is inserted into Eq.~(\ref{eq:usualtraceanom}), the trace anomaly is not one-loop exact.
Then it is inconsistent with the Adler-Bardeen theorem and the supercurrent equation.

However, the beta function is obviously renormalization scheme dependent. 
For example, the beta function in the so-called DR (or $\overline{\rm DR}$) scheme is not given by the NSVZ beta function 
beyond the two-loop level~\cite{Jack:1996cn}.
We can even define a gauge coupling, called the holomorphic gauge coupling,
whose beta function is one-loop exact~\cite{Shifman:1986zi}. 
Furthermore, the composite operator $F_{\m\n}F^{\m\n}$ is also renormalization scheme
dependent. Eq.~(\ref{eq:usualtraceanom}) has been proven~\cite{Adler:1976zt,Collins:1976yq,Nielsen:1977sy} by using
the minimal subtraction scheme in dimensional regularization for QCD-like theories, but this scheme cannot preserve 
manifest supersymmetry. Because the anomaly puzzle originates from the superspace equation,
we have to use a scheme in which supersymmetry is preserved in a manifest way. 

In this paper, we perform calculations of the trace anomaly by employing the ${\cal N}=4$ regularization of the ${\cal N}=1$ 
pure Yang-Mills~\cite{ArkaniHamed:1997mj,Dine:2011gd},
which is suitable for our purpose. 
(Regularization by dimensional reduction (DRED)~\cite{Siegel:1979wq,Capper:1979ns} can also maintain manifest supersymmetry.\footnote{We
will not care about mathematical inconsistency, violation of supersymmetry, and problems with $\g_5$ 
in DRED. See Ref.~\cite{Stockinger:2005gx}
and references therein for discussions on these things. 
}
But this scheme is 
not suitable for the study of the anomaly puzzle, as we explain in section~\ref{sec:5}.)
We will find that the trace anomaly is indeed given by the one-loop exact form, if the operator $F_{\m\n}F^{\m\n}$
is renormalized in a way which is preferred by supersymmetry and the topological nature of $F_{\m\n}\tilde{F}^{\m\n}$, where 
$\tilde{F}^{\m\n}=\frac{1}{2}\e^{\m\n\r\s}F_{\m\n}$.

In section~\ref{sec:2}, we give an indirect derivation of the trace anomaly in general field theories,
and clarify the point at which the scheme dependence enters. In section~\ref{sec:3} the ${\cal N}=4$ regularization is introduced.
In section~\ref{sec:4}, the trace anomaly formula is applied to the ${\cal N}=4$ regularized version of the ${\cal N}=1$ theory.
There we discuss how the scheme dependence becomes important. Finally, conclusions are given in section~\ref{sec:5}.
We also discuss the relation between our work and 
the previous works~\cite{Grisaru:1985yk,Grisaru:1985ik,Shifman:1986zi,ArkaniHamed:1997mj}.
\ref{app:A} contains a review of basic properties of the energy-momentum tensor.

\section{Trace anomaly formula}
\label{sec:2}
In this section, we give an indirect derivation of the trace anomaly (up to a total derivative term) in a general field theory which has a Lagrangian description.
We will clarify the assumption which is made in the derivation of the trace anomaly.

Let us consider a theory described by a set of fields $\v_I$, where $I$ is a label specifying the fields. 
We assume that the theory is renormalized in some renormalization scheme.
We consider a correlation function,
\beq
\vev{\prod_i \v_{I_i}(x_i)}=\int [D\v] \exp(iS[\v,\l_a])\prod_i \v_{I_i}(x_i),
\eeq
where $\l_a$ are parameters (e.g., couplings and masses) of the theory, and $S$ is the action. 
We assume that the field $\v_I$ has mass dimension $D_I$ and the parameter $\l_a$ has mass dimension $d_a$.
By a simple dimensional analysis, we obtain
\beq
0=\left(- \sum_i \left(x_i^\m \frac{\q}{\q x^\m_i} +D_{I_i} \right) +\m \frac{\q}{\q \m}+ \sum_a d_a \l_a \frac{\q}{\q \l_a}   \right) \vev{\prod_i \v_{I_i}(x_i)}, 
\label{eq:dimanalysis}
\eeq
where $\mu$ is the renormalization scale.
On the other hand, the renormalization group (RG) equation tells us that
\beq
0=\left( \m \frac{\q}{\q \m}+ \sum_a \b_a \frac{\q}{\q \l_a} +\sum _i \g_{I_i}  \right) \vev{\prod_i \v_{I_i}(x_i)}.
\label{eq:rganalysis}
\eeq
where $\g_I$ is the anomalous dimension of $\v_I$, and $\b_a$ is the beta function defined by
\beq
\b_a = \m \frac{\q}{\q \m} \l_a.
\eeq
We assume that there are no operator mixings among $\v_I$ for simplicity.

Next, by using the Ward-Takahashi identity of the energy-momentum tensor given by Eq.~(\ref{eq:WTidentityofEMtensor}) of \ref{app:A},
we obtain
\beq
\vev{\int d^dy iT^\m_\m(y)\prod_i \v_{I_i}(x_i)}=
\sum_i 
 \left( x_i^\m  \frac{\q}{\q x^\m_i}+d C_{I_i} \right)  \vev{\prod_{ i} \v_{I_i}(x_i)}. \label{eq:tracewt}
\eeq
where $d$ is the space-time dimension, and $C_I$ are arbitrary constants representing the ambiguity in the definition of the energy-momentum tensor.
(see \ref{app:A} for details).
We have assumed that the integral of the total derivative $\int d^d x \q_\m(x_\n T^{\m\n})$ vanishes.\footnote{
We should subtract the ``cosmological constant term'' in $T^{\m\n}$ so that the vacuum expectation value is given by $\vev{T^{\m\n}}=0$.}
Combining Eqs.~(\ref{eq:dimanalysis}), (\ref{eq:rganalysis}) and (\ref{eq:tracewt}), we obtain
\beq
\vev{\int d^dy iT^\m_\m(y)\prod_i \v_{I_i}(x_i)}=
\left( \sum_i \left(d C_{I_i}-\D_{I_i} \right) + \sum_a (d_a \l_a-\beta_a) \frac{\q}{\q \l_a}   \right) \vev{\prod_i \v_{I_i}(x_i)},
\eeq
where we have defined $\D_I=D_I+\g_I$.
We can use Eq.~(\ref{eq:numbercount}) to rewrite the first term of the right-hand-side as
\beq
\left( \sum_i \left(d C_{I_i}-\D_{I_i} \right) \right) \vev{\prod_i \v_{I_i}(x_i)}=
-i\vev{\left( \int d^d y \sum_I \left(d C_{I}-\D_{I} \right)N_{I}(y) \right) \prod_i \v_{I_i}(x_i)},
\eeq
where $N_I$ is an operator which vanishes by equations of motion (see \ref{app:A}).

Now we make a crucial assumption. Naively, by using the path integral expression for the correlation function,
the derivative with respect to $\l_a$ is given as
\beq
\frac{\q}{\q \l_a}\vev{  \prod_i \v_{I_i}(x_i) }  =\vev{i\int d^d y \frac{\q {\cal L}(y)}{\q \l_a}     \prod_i \v_{I_i}(x_i)    },\label{eq:actionprinciple}
\eeq
where ${\cal L}$ is the Lagrangian. 
This is known as the quantum action principle~\cite{Lowenstein:1971jk,Breitenlohner:1977hr}. 
Assuming that this is really the case, we obtain 
\beq
&&\vev{\int d^dy T^\m_\m(y)\prod_i \v_{I_i}(x_i)}= \nonumber \\
&&\vev{ \int d^d y \left( \sum_I  \left( \D_{I}-d C_{I} \right)  N_{I}(y)  + \sum_a (d_a \l_a-\beta_a) \frac{\q {\cal L}(y)}{\q \l_a} \right)   \prod_i \v_{I_i}(x_i) }.
\eeq
Therefore, we finally obtain the operator equation for the trace anomaly,
\beq
T^\m_\m=  \sum_a (d_a \l_a-\beta_a) \frac{\q {\cal L}}{\q \l_a}+\sum_I  \left( \D_{I}-d C_{I} \right)  N_{I} +\q_\m j^\m_V, \label{eq:traceanomalyformula}
\eeq
where $\q_\m j^\m_V$ is a total derivative term which we do not try to determine. As noted above, $N_I$ vanishes by equations of motion.
Therefore, as expected, $T^\m_\m $ vanishes up to the total derivative term in the fixed point of RG flow, $d_a \l_a-\beta_a=0$.

In the above derivation, the use of the quantum action principle~(\ref{eq:actionprinciple}) is crucially important.
The validity of the quantum action principle depends on what regularization or renormalization 
scheme is used. For example, it is known that Eq.~(\ref{eq:actionprinciple}) holds when minimal subtraction in dimensional regularization is used
to renormalize all the parameters and operators of the theory~\cite{Breitenlohner:1977hr}. Then, Eq.~(\ref{eq:traceanomalyformula}) 
reproduces the result of Refs.~\cite{Adler:1976zt,Collins:1976yq,Nielsen:1977sy} for the trace anomaly in QCD-like theories up to a total derivative term. 

If Eq.~(\ref{eq:actionprinciple}) is naively applied in the ${\cal N}=1$ pure-Yang-Mills, we would obtain
\beq
\frac{\q}{\q g^{-2}} \vev{ (\cdots)} = i\vev{\left(\int d^4x d^2\h\frac{1}{4}W^\a W_\a+{\rm h.c.}\right)(\cdots)} \label{eq:naiveaction}
\eeq
where $(\cdots)$ represents a chain of operators. If this is the case, the trace anomaly is proportional to $(\b/g^4) W^\a W_\a$.
In the following sections, we examine in what scheme Eq.~(\ref{eq:naiveaction}) is really valid in the ${\cal N}=1$ pure Yang-Mills.

\section{${\cal N}=4$ regularization of ${\cal N}=1$ pure Yang-Mills}
\label{sec:3}

We employ the ${\cal N}=4$ regularization of the ${\cal N}=1$ pure Yang-Mills which is discussed in Refs.~\cite{ArkaniHamed:1997mj,Dine:2011gd}.
In this regularization, we take the ${\cal N}=4$ theory with ${\cal N}=4$ supersymmetry softly broken to ${\cal N}=1$ by the mass terms
of the adjoint chiral fields. The Lagrangian is given by~\footnote{
We follow the notation and convention of Wess and Bagger~\cite{Wess:1992cp},
except that a matter kinetic term is given by $\F^\dagger e^{-2V}\F$, and a gauge field strength chiral field is given by
$W_\a=\frac{1}{8}\bar{D}^2 (e^{2V}D_\a e^{-2V})$.
}
\beq
{\cal L} &=&\int d^2\h \frac{1}{4g^2_h}W^{\a} W_\a +{\rm h.c.} +  \int d^2\h d^2\bar{\h} \sum_{i=1}^3 Z_i \F_i^\dagger e^{-2V} \F_i  \nonumber \\
&&+\int d^2 \h \left( \sqrt{2} f_{ABC}\F^A_1\F^B_2\F^C_3 +\sum_{i=1}^3\frac{1}{2}M_i \F_i^2 \right)+{\rm h.c.},
\eeq
where $A,B$ and $C$ are gauge indices, $f_{ABC}$ is the structure constant of the gauge group,
and gauge indices are omitted in other terms. The parameters $g_h^2$ and $Z_i~(i=1,2,3)$ are given in terms of the gauge coupling $g$
and the theta angle $\h$ as
\beq
g^{-2}_h=g^{-2}-\frac{i \h}{8\pi^2}, \\
Z_1Z_2Z_3={\rm Re} (g^{-2}_h), \label{eq:constwavefunction}
\eeq
where the second equation comes from the constraint of ${\cal N}=4$ supersymmetry.
Following Ref.~\cite{Dine:2011gd}, we have normalized the fields so that the coefficient of the term $ f_{ABC}\F^A_1\F^B_2\F^C_3$ does not depend on the gauge coupling $g$.
In this normalization, we can maintain the holomorphy~\cite{Seiberg:1993vc} when the holomorphic coupling $g_h$ and the mass terms $M_i~(i=1,2,3)$
are extended to background chiral superfields.

In the energy region much below the scale $M_i$, this theory becomes the same as the ${\cal N}=1$ pure Yang-Mills, up to terms
which are power suppressed by $M_i$. However, the existence of the chiral multiplets $\F_i$ makes the theory UV-finite thanks to the known finiteness of 
the ${\cal N}=4$ theory.\footnote{It is important that the ${\cal N}=4$ supersymmetry is only softly broken by the mass terms of the 
adjoint chiral fields, and hence the mass terms do not affect the UV divergences of dimensionless quantities. }
Therefore, this theory can be regarded as the ${\cal N}=1$ pure Yang-Mills regulated by the adjoint chiral fields with the cutoff scale $M_i$.

Before closing this section, some technical remarks are in order.
Just to make our arguments more concrete, we further regulate the theory by e.g., regularization by dimensional reduction (DRED).
One reason that we introduce the further regularization is that we would like the regularization to be consistent with the extension of parameters
into background superfields so that we can utilize the power of holomorphy. Strictly speaking, this extension maintains only
${\cal N}=1$ supersymmetry, and a possibility exists that the ${\cal N}=4$ regularization might not be compatible with the extension of couplings into 
background superfields.
However, we believe that there is no such incompatibility.  
For example, let us suppose that 
the couplings $g^{-2}_h$ and $Z_i$ (and also $M_i$) are allowed to have non-vanishing constant $F$-terms (and $D$-terms for $Z_i$) with the constraint 
(\ref{eq:constwavefunction}) imposed as a superfield equation. 
Then, the violation of ${\cal N}=4$ supersymmetry is only soft, i.e., it is broken only by parameters with 
positive mass dimensions. Therefore all the divergences in dimensionless quantities are still cancelled, and we need no counterterms
for the lowest components of $g^{-2}_h$ and $Z_i$. Other soft breaking parameters are contained in the same multiplet as $g^{-2}_h$ or $Z_i$,
so it is expected that no counterterms are required also for these soft breaking parameters. 
This idea may be elegantly realized in the scheme of analytic continuation of parameters into superspace discussed in 
Ref.~\cite{ArkaniHamed:1998kj}. For example, RG equations have been given for soft breaking terms
by analytically continuing the RG equations for $g_h^{-2}$ and $Z_i$ into superspace, and hence all the soft breaking terms are
manifestly RG invariant if that is the case for the lowest components. This suggests that all the UV divergences are cancelled 
in the present theory even if the parameters are analytically continued as in Ref.~\cite{ArkaniHamed:1998kj}.\footnote{
Here we have introduced the framework of Ref.~\cite{ArkaniHamed:1998kj} not because we want to have soft breaking parameters in our theory,
but because we want to make the statements about holomorphy more explicit. For our purposes, it is important that we are working in a well-defined
(i.e., explicitly regularized) setting where the argument of holomorphy can (at least in principle) be made rigorous. }

An important check of the cancellation of divergences in our regularization framework 
can be found in Ref.~\cite{ArkaniHamed:1998kj}. When $g^{-2}_h$ and $Z_i$ have nontrivial soft terms,
a counterterm for the mass of the so-called epsilon scalar in DRED is required. 
At the one-loop level, this counterterm, when analytically continued into superspace,
is given as
\beq
-\int d^2\h d^2\bar{\h}  \frac{1}{16 \pi^2} \frac{1}{\e}   \left[ t_{ Ad} \log ({\rm Re}(g_h^{-2}))-\sum_i t_i \log Z_i  \right] (\G^{\hat{\hat{\m}}}\G_{\hat{\hat{\m}}}),
\label{eq:epsilonscalarcounterterm}
\eeq
where $\hat{\hat{\m}}$ is the Lorentz index of the ``compactified'' $2\e$ dimensions in DRED, $t_{Ad}$ and $t_i$ are Dinkin indices for
the adjoint representation and the representation of the matter field $i$ respectively,
and $\G_\m$ is defined as
\beq
\G_\m = \frac{1}{4} \bar{\s}_\m^{\dot{\a}\a}\bar{D}_{\dot{\a}}(e^{2V}D_\a e^{-2V}).
\eeq
One can check that $\G_{\hat{\hat{\m}}}$ is gauge covariant, and hence the above term is allowed by gauge invariance~\cite{Grisaru:1985tc}.
If Eq.~(\ref{eq:epsilonscalarcounterterm}) were present, this would give a contribution to the gauge kinetic term by using 
$\bar{D}^2  (\G^{\hat{\hat{\m}}}\G_{\hat{\hat{\m}}})=4\e W^\a W_\a $.
In our theory, we have $t_i=t_{Ad}~(i=1,2,3)$, so Eq.~(\ref{eq:epsilonscalarcounterterm}) vanishes thanks to the condition (\ref{eq:constwavefunction}).
For this cancellation to happen, it is important that the adjoint chiral fields are normalized in the way which preserves 
the holomorphy of the superpotential~\cite{Dine:2011gd}, leading to Eq.~(\ref{eq:constwavefunction}).
We expect that this cancellation continues to happen at higher order level. In the following, we assume that
the ${\cal N}=4$ regularization is compatible with the extension of the parameters to ${\cal N}=1$ 
superfields (at least if they are space-time constants, i.e., $\q_\m g_h^{-2}=0$ etc.), and the usual argument
based on holomorphy is justified in our regularization framework.\footnote{Much easier way to ensure finiteness may be to use ${\cal N}=2$
supersymmetry, by taking $Z_1={\rm Re}(g^{-2}_h)$ and $Z_2=Z_3=1$. In this case, we can extend $g^{-2}_h$ to a background ${\cal N}=2$
vector multiplet~\cite{Argyres:1996eh}. Then the absence of radiative corrections is established. }

Another reason we regularize the ${\cal N}=4$ theory by e.g. DRED 
is that we need to treat composite operators. Even in ${\cal N}=4$
theories, composite operators require regularization and renormalization in general. However, fortunately, this problem
is also not essential for our purpose of computing the trace anomaly. In our analysis, the only composite operators
which we treat are the energy-momentum tensor
and the operators appearing in the trace anomaly.
The energy-momentum tensor is finite
(at least if improvement~\cite{Callan:1970ze} is done appropriately).
On the other hand, we need the operator $\q {\cal L}/\q g^{-2}$ for the trace anomaly.
For simplicity, let us neglect the mass terms for the adjoint chiral fields.
Then this operator is indeed finite in ${\cal N}=4$ theories. The most easy way to see this is to note that 
the coupling $g$ is exactly marginal, i.e., the coupling does not run in the RG flow for all the possible values of it. 
Then the operator $\q {\cal L}/\q g^{-2}$ should have scaling dimension 4 (at least up to possible total derivative terms
which we do not care), which coincides with its mass dimension.
Therefore there is no anomalous dimension for this operator. This fact indicates that no renormalization is required for this operator.

\section{The trace anomaly in ${\cal N}=1$ pure Yang-Mills} 
\label{sec:4}
Now we discuss the calculations of the beta functions and the trace anomaly in the ${\cal N}=1$ pure Yang-Mills.
Let us first discuss the beta functions. In the previous section, we have argued that the manifest holomorphy
is expected to be maintained in the regularization process.
Then, the low energy physics depends only on the combination 
\beq
\frac{8\pi^2}{g_h^2}-\sum_{i=1}^3 t_{Ad}\log M_i. \label{eq:lowenergycoupling}
\eeq
This statement can be confirmed directly at the one-loop level (by decoupling the adjoint chiral fields, 
perhaps in the manifestly supersymmetric framework of Ref.~\cite{ArkaniHamed:1998kj}). 
Higher order corrections are absent due to the holomorphy and
the dependence of $g_h^{-2}$ on the theta angle.\footnote{This argument for the one-loop exactness 
becomes invalid when we try to apply it for SQCD regularized by
finite ${\cal N}=2$ theories. In the low-energy theory, we have to regard matter wave-function renormalization as an independent parameter in addition to the
holomorphic coupling, as is evident in the analyses of e.g. Ref.~\cite{ArkaniHamed:1998kj}. We have two (or more) parameters in the low-energy theory,
but there is only one gauge coupling $g$ in the high-energy theory. Then it is not possible to choose $g$ so as to fix both the holomorphic gauge coupling
and the wave-function renormalization of the low-energy theory when the cutoff scale is changed. This difficulty can be avoided by
rescaling the matter fields so that the wave function renormalization of the low energy theory becomes unity. 
However, this process violates the holomorphy, and hence
the one-loop exactness is lost.}

We introduce a single cutoff scale $M$ (which is taken to be a real positive number) and will give the masses $M_i$ in terms of $M$.
We have infinitely many choices for doing this~\cite{Dine:2011gd}, and this choice determines the renormalization prescription for the
bare coupling $g$. We will study the RG equation for $g$ which makes the combination (\ref{eq:lowenergycoupling})
invariant when the cutoff scale $M$ is changed. Although we discuss the RG equation for the bare quantity, it is also easy to define renormalized couplings
in our regularization framework. We will comment on this point later. 

Two particular choices for $M_i$ are often discussed.
One choice is given by $M_i=M$, which maintains the holomorphy about the cutoff $M$.
In this case, by requiring that the combination (\ref{eq:lowenergycoupling})
is invariant under the change of $M$, we obtain
\beq
M\frac{\q}{\q M} g^{2}=-\frac{3t_{Ad}g^4}{8\pi^2}.
\eeq
Thus we obtain the one-loop exact beta function in this scheme. The other choice is that we take $M_i=Z_i M$,
so that the cutoff $M$ corresponds to the tree level masses of $\F_i$ when these fields are canonically normalized. 
Then, the real part of Eq.~(\ref{eq:lowenergycoupling}) becomes
\beq
\frac{8\pi^2}{g^2}-\sum_{i=1}^3 t_{Ad}\log (Z_i M)=\frac{8\pi^2}{g^2}-t_{Ad}\log g^{-2} -3 t_{Ad} \log (M)
\eeq
where Eq.~(\ref{eq:constwavefunction}) has been used. This choice leads to the NSVZ beta function
\beq
M\frac{\q}{\q M} g^{2}=-\frac{1}{8\pi^2}\frac{3t_{Ad}g^4}{1-t_{Ad}g^2/8\pi^2}. \label{eq:NSVZbetafunciton}
\eeq

\begin{figure}
\begin{center}
\includegraphics[scale=0.3]{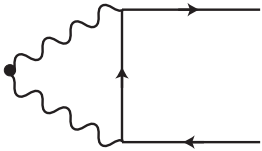}
\caption{An example of the divergence of the operator $F_{\m\n} \tilde{F}^{\m\n}$. Wavy lines are gauge fields, and solid lines are fermions.
The black filled circle is the operator $F_{\m\n} \tilde{F}^{\m\n}$.}
\label{fig:1}
\end{center}
\end{figure}

Now we would like to discuss the trace anomaly. Before doing that, a remark on the operator $W^\a W_\a$ 
is necessary. The trace anomaly is given in terms of the composite operator $F_{\m\n}F^{\m\n}$,
which is contained in $W^\a W_\a$.
This superfield also contains the operator $F_{\m\n}\tilde{F}^{\m\n}$.
Due to its topological property,
the operator $F_{\m\n}\tilde{F}^{\m\n}$ should not receive multiplicative renormalization, and hence
the operator $W^\a W_\a$ also should have a preferred overall scale. (However, see section~\ref{sec:5} for discussion on trouble in DRED 
if the ${\cal N}=4$ regularization is not used.)
This does not mean that the operator $F_{\m\n}\tilde{F}^{\m\n}$ requires no renormalization at all~\cite{Espriu:1982bw,Breitenlohner:1983pi}.
Although the overall scale of $F_{\m\n}\tilde{F}^{\m\n}$ should be determined by its topological nature, radiative corrections
of the form $F_{\m\n}\tilde{F}^{\m\n} \to F_{\m\n}\tilde{F}^{\m\n}+\q_\m j^\m$ is not restricted by topological arguments, where $j_\m$
is a gauge invariant operator. In the language of differential forms, we can add an exact 4-form $d(* j)$ to the closed 4-form $F \wedge F$ without affecting
it as an element of the de~Rham cohomology group. For example, in the case of QED, the one-loop diagram in Figure~\ref{fig:1} gives a divergence and hence
the operator $F_{\m\n}\tilde{F}^{\m\n}$ needs a counterterm given by
\beq
[F_{\m\n}\tilde{F}^{\m\n}] = F_{\m\n}\tilde{F}^{\m\n}+\left( \frac{3e^4 \m^{2\e}}{8\pi^2} \frac{1}{\e} +c_{\rm finite} \right)\q_\m j_5^\m,
\eeq
where we have used dimensional regularization,\footnote{Here we have avoided DRED 
because DRED is extremely subtle to perform this type of calculations, because of the 
relation $\tr(\g_5 \g^\m \g^\n \g^\r \g^\s)=0$ (see e.g. Ref~\cite{Stockinger:2005gx}). 
We hope that this $\g_5$ problem does not cause trouble as long as we are working in superspace without looking components of superfields.} 
$j_5^\m=\frac{i}{2}\bar{\p}[\g^\m,\g_5]\p$ is the axial current,
and $[F_{\m\n}\tilde{F}^{\m\n}]$ is the renormalized operator. 
The term $c_{\rm finite}$ is a finite counterterm whose value depends on renormalization prescriptions,
and hence the operator $[F_{\m\n}\tilde{F}^{\m\n}]$ has an ambiguity of the form $c_{\rm finite}\q_\mu j^\m_5$.

One might think that the distinction between $F_{\m\n}\tilde{F}^{\m\n}$ and $\q_\m j^\m_5$ is meaningless (in massless limit of fermions)
because the axial anomaly equates them. If that were the case, $F_{\m\n}\tilde{F}^{\m\n}$ would get multiplicative renormalizations. 
However, the operator $\q_\m j^\m_5$ is equivalent to $F_{\m\n}\tilde{F}^{\m\n}$
only if we use equations of motion, i.e., $\q_\m j^\m_5 \approx F_{\m\n}\tilde{F}^{\m\n}/8\pi^2$, 
and they differ by the presence of contact terms when they are inserted into correlation functions.
The contact terms are essential in the discussion of the topological properties of $F_{\m\n}\tilde{F}^{\m\n}$.
Meaning of the Adler-Bardeen theorem also requires careful considerations.
See Ref.~\cite{Yonekura:2010mc} for discussions on the importance of this observation in solving the anomaly puzzle.

In supersymmetric theories, the radiative corrections and counterterms discussed above are extended to superspace as
\beq
[W^\a W_\a] = W^\a W_\a +\bar{D}^2 J,
\eeq
where $J$ is a real vector superfield. Therefore this type of corrections should be allowed in the definition of the composite operator $W^\a W_\a$.

Let us calculate the trace anomaly using the formula~(\ref{eq:traceanomalyformula}). In the previous section,
we have regularized the theory further by DRED for concreteness. It is known that the quantum action principle~(\ref{eq:actionprinciple}) holds
for bare quantities in DRED~\cite{Stockinger:2005gx}. Thus we can use Eq.~(\ref{eq:actionprinciple}) in our regularization.
The important point is that we have to include regulator contributions which come from terms containing $\F_i$.

If we use the scheme $M_i=M$ which have led to the one-loop exact beta function, we obtain~\footnote{
We also have to include contributions from gauge fixing terms and ghosts. However, we expect that these contributions are BRST exact,
as in the case of QCD~\cite{Adler:1976zt,Collins:1976yq,Nielsen:1977sy}. We neglect them in this paper.}
\beq
\frac{\q {\cal L}}{\q g^{-2}} 
&=& \int d^2\h \frac{1}{4}\left(W^{\a} W_\a +{\bar D}^2 K \right)+{\rm h.c.} \\
K&=&-\frac{1}{2}\sum_{i=1}^3 \frac{\q Z_i}{\q g^{-2}} \F_i^\dagger e^{-2V} \F_i. \label{eq:wwcounterterm}
\eeq
As explained in the previous section, the operator ${\q {\cal L}}/\q g^{-2}$ is finite due to ${\cal N}=4$ supersymmetry. 
However, the operator $W^{\a} W_\a$ itself receives divergent corrections even in ${\cal N}=4$ theories
from diagrams such as Figure~\ref{fig:1}. Thus the term $\bar{D}^2 K$
can be regarded as a regulator term which makes the operator $W^\a W_\a$ finite. We define the renormalized operator as~\footnote{
The operator $K$, and hence $[W^\a W_\a]_R$, may look to be depending on the precise choice of $Z_i$, but it is not so. 
If we use different wave function renormalizations $Z'_i$ satisfying Eq.~(\ref{eq:constwavefunction}) 
and obtain $K'$, then the difference $K'-K$ is a current of $\SU(3) \subset \SU(4)_R $ symmetry of the ${\cal N}=4$ theory.
In particular, it is UV-finite. Then the argument of the heavy field decoupling theorem (see e.g. Ref.~\cite{Collins:1984xc}) 
may tell us that $K'-K$ is power suppressed
by the heavy field masses $M_i$. The UV-finiteness is essential, because otherwise momentum integrals such as 
$\int d^4 k (k^2+M^2)^{-2}$ spoil the suppression.
}
\beq
[W^\a W_\a]_R = W^{\a} W_\a +{\bar D}^2 K .
\eeq
As explained above, the addition of ${\bar D}^2 K$ does not violate the topological nature of $F_{\m\n}\tilde{F}^{\m\n}$, and hence it should be allowed.
Therefore, the trace anomaly is given as
\beq
T^\m_\m &\approx& -\frac{\q g^{-2}}{\q M} \frac{\q {\cal L}}{\q g^{-2}} + \q_\m j^\m_V \nonumber \\
&=& - \frac{3t_{Ad}}{32 \pi^2}  \int d^2\h [W^{\a} W_\a]_R+{\rm h.c.} +\q_\m j^\m_V \label{eq:N=1traceanomaly}
\eeq
where $\approx$ means that we have used the equations of motion, $N_I \approx 0$.
Therefore, we have determined that the coefficient of the operator $[W^\a W_\a]_R$ in the trace anomaly is one-loop exact.

Next, let us perform the calculation in the scheme $M_i=Z_i M$, which have led to the NSVZ beta function. Then, we obtain
\beq
\frac{\q {\cal L}}{\q g^{-2}} 
= \int d^2\h \left( \frac{1}{4} [W^{\a} W_\a]_R +\frac{1}{2} \sum_{i=1}^3 \frac{\q  \log Z_i}{\q g^{-2}}  M_i \F_i^2 \right)+{\rm h.c.} .
\eeq
In this case, we have additional contributions coming from the regulator mass terms. 
Therefore, the naive use of Eq.~(\ref{eq:naiveaction}) is invalid if the regulator contribution is not taken into account seriously.
In the low energy limit (i.e., in the energy region much below $M$), the operator $M_i \F_i^2$ becomes 
\beq
M_i \F_i^2 \to -\frac{t_{Ad}}{16 \pi^2}[W^{\a} W_\a]_R~~~~(i=1,2,3). \label{eq:operatorflow}
\eeq
At the one-loop level, this equation can be checked directly.
The holomorphy restricts higher order corrections to the form $W^\a W_\a \to W^\a W_\a + {\bar D}^2 J$.
We have assumed that these corrections are such that the operator $W^\a W_\a$ is replaced by $[W^\a W_\a]_R$. This is because
the operator $\F_i^2$ is UV-finite since it is protected by ${\cal N}=4$ superconformal symmetry.

Using Eq.~(\ref{eq:operatorflow}), we obtain
\beq
\frac{\q {\cal L}}{\q g^{-2}} 
&\to&  \int d^2\h  \frac{1}{4} \left(1- \frac{t_{Ad}}{8 \pi^2} \sum_{i=1}^3 \frac{\q  \log Z_i}{\q g^{-2}}\right) [W^{\a} W_\a]_R    +{\rm h.c.}  \nonumber \\
&=&   \int d^2\h  \frac{1}{4} \left(1- \frac{t_{Ad} g^2 }{8 \pi^2} \right) [W^{\a} W_\a]_R    +{\rm h.c.} ,\label{eq:principleviolated}
\eeq
where we have used Eq.~(\ref{eq:constwavefunction}).
Combining this result with the NSVZ beta function~(\ref{eq:NSVZbetafunciton}), we again get the same answer given by Eq.~(\ref{eq:N=1traceanomaly}).
Therefore, the one-loop exact form of the trace anomaly is also confirmed in this scheme.

Finally, let us comment on renormalization in the ${\cal N}=4$ regularization. We have treated bare couplings up to now, 
but it is very easy to define a renormalized coupling since we know that the low energy physics depends only on the combination~(\ref{eq:lowenergycoupling}).
For example, we can define a renormalized coupling $g_R$ as
\beq
g_R^{-2}(\mu)=g^{-2}-\sum_{i=1}^3 \frac{t_{Ad}}{8\pi^2}\log \left( \frac{M_i}{\m} \right),
\eeq
where $\m$ is the renormalization scale. Now we can let the cutoff $M_i$ be infinity at the end of the calculations with $g_R$ fixed.
The derivation of the trace anomaly is completely parallel to the one given above.
Although it is not so evident that the operator $[W^\a W_\a]_R$ remains finite in the limit $M_i \to \infty$, we expect that this is the case
since the energy-momentum tensor (and hence its trace) is UV-finite.

\section{Conclusions and comparison with the literature}
\label{sec:5}
In this paper we have investigated the trace anomaly in ${\cal N}=1$ pure Yang-Mills to solve the anomaly puzzle in this theory.
The trace anomaly is often quoted as $-(\b/4g^4)F_{\m\n}F^{\m\n}$, but the validity of this formula depends on
whether the quantum action principle given by Eq.~(\ref{eq:naiveaction}) (or more generally Eq.~(\ref{eq:actionprinciple})) is really true or not.
To settle this issue, we have explicitly regularized the theory by using the ${\cal N}=4$ regularization. 
Then the regulator contribution to Eq.~(\ref{eq:naiveaction}) can be explicitly seen, and this contribution depends on what scheme we use.
In the scheme in which we obtain the one-loop exact beta function, it is almost correct 
to use Eq.~(\ref{eq:naiveaction}) naively 
with the regulator contribution neglected.
The regulator contribution only plays a role of making the operator $W^\a W_\a$ renormalized, $[W^\a W_\a]_R=W^\a W_\a+{\bar D}^2K$, without 
spoiling the topological nature of $F_{\m\n}\tilde{F}^{\m\n}$ contained in $W^\a W_\a$.
However, if the scheme is used  in which we obtain the NSVZ beta function,
the regulator contribution is not at all negligible, as seen in Eq.~(\ref{eq:principleviolated}). 
If this contribution is correctly taken into account, we obtain the same answer for the trace anomaly in both of the schemes.
The result for the trace anomaly has the one-loop exact form, and hence is consistent with the supercurrent equation and the Adler-Bardeen theorem.

Although the ${\cal N}=4$ regularization discussed in this paper is applicable only to the ${\cal N}=1$ pure Yang-Mills,
we strongly believe that there exists a renormalization scheme in which the supercurrent equations of Refs.~\cite{Shifman:1986zi,Yonekura:2010mc}
which include matter fields are really justified. This ``desirable renormalization scheme'' should also justify various exact results in supersymmetric gauge theories.
One of the properties which the desirable scheme should possess is the quantum action principle~(\ref{eq:actionprinciple}),
at least if we use the couplings which can be extended to background superfields.

In the rest of this section, we would like to comment on how our result is consistent with the results obtained in 
Refs.~\cite{Grisaru:1985yk,Grisaru:1985ik,Shifman:1986zi,ArkaniHamed:1997mj}.
Grisaru, Milewski and Zanon~\cite{Grisaru:1985yk,Grisaru:1985ik} studied the supercurrent equation by employing DRED.
They constructed two supercurrents which we will call ${\cal J}^{(1)}_{\a\dot{\a}}$ and ${\cal J}^{(2)}_{\a\dot{\a}}$, and performed 
calculations at the two-loop level.
One supercurrent ${\cal J}^{(1)}_{\a\dot{\a}}$ satisfies the equation (in our notation and convention)
\beq
\bar{D}^{\dot{\a}} {\cal J}^{(1)}_{\a\dot{\a}} \approx D_\a \left( \frac{1}{3} \frac{\b_{\rm two-loop}}{g^4}[W^\a W_\a]_{\rm DR} \right), \label{eq:gmsone}
\eeq
where the coupling $g$ and the operator $[W^\a W_\a]_{\rm DR}$ are renormalized by minimal subtraction in DRED (i.e., the DR or 
$\overline{\rm DR}$ scheme),
and $\b_{\rm two-loop}$ is the two-loop beta function which coincides with the NSVZ beta function at the two-loop level.
The other supercurrent ${\cal J}^{(2)}_{\a\dot{\a}}$ satisfies
\beq
\bar{D}^\a {\cal J}^{(2)}_{\a\dot{\a}} \approx  D_\a \left( \frac{1}{3} \frac{\b_{\rm one-loop}}{g^4}[W^\a W_\a]_{\rm DR} \right)+
b\bar{D}^2 D_\a[\G^{\hat{\hat{\m}}}\G_{\hat{\hat{\m}}}]_{\rm DR} ,\label{eq:gmstwo}
\eeq
where $\b_{\rm one-loop}$ is the one-loop beta function, and $b$ is a coefficient which they did not explicitly compute. 
They claimed that ${\cal J}^{(1)}_{\a\dot{\a}}$ contains the energy-momentum tensor, while the lowest component of ${\cal J}^{(2)}_{\a\dot{\a}}$
satisfies the Adler-Bardeen theorem.

Let us first consider Eq.~(\ref{eq:gmsone}). At first sight, this equation may look inconsistent with our result, because 
it seems to give the trace anomaly which is not one-loop exact.
However, this is not necessarily the case. The problem is that the normalization of the operator $W^\a W_\a$ is quite ambiguous in DRED.
We have discussed in section~\ref{sec:4} that this operator has an ambiguity of adding a term $\bar{D}^2 J$, where $J$ is a gauge invariant 
real vector superfield.
However, in DRED, there exists the operator $\G^{\hat{\hat{\m}}}\G_{\hat{\hat{\m}}}$ (see section~\ref{sec:3}) satisfying 
\beq
\bar{D}^2  (\G^{\hat{\hat{\m}}}\G_{\hat{\hat{\m}}})=4\e W^\a W_\a . \label{eq:wwtotder}
\eeq
By choosing $J \propto (\G^{\hat{\hat{\m}}}\G_{\hat{\hat{\m}}})$, the normalization of the operator $W^\a W_\a$ becomes completely ambiguous.
This ambiguity may be related to the fact that there is no instanton in $4-2\e$ dimensions, and hence the topological arguments do not work in this scheme.

More concretely, suppose that the divergences of $W^\a W_\a$ are of the form $(1/\e^n) \bar{D}^2 (\G^{\hat{\hat{\m}}}\G_{\hat{\hat{\m}}})$
$(n=1,2,\cdots)$~\cite{Grisaru:1985tc}, and
we define a renormalized operator $[W^\a W_\a]$ as
\beq
[W^\a W_\a] &=& W^\a W_\a +\sum_{n=1}^\infty \frac{c_n(g)}{\e^n} \frac{1}{4}\bar{D}^2 (\G^{\hat{\hat{\m}}}\G_{\hat{\hat{\m}}}) \nonumber \\
&=&  \left(1+c_1(g)  \right) W^\a W_\a+ \sum_{n=2}^\infty \frac{c_n(g)}{\e^{n-1}} W^\a W_\a .
\eeq
The $n=1$ term eventually gives a finite counterterm.
Then, we have a choice for the renormalization prescription. One choice is to set $c_1(g)=0$, which corresponds to 
the DR scheme. But we can also choose a scheme in which $c_1(g) \neq 0$. The freedom in choosing $c_1(g)$ leads to the
ambiguity of the renormalized operator $[W^\a W_\a]$. 

One may consider that it is meaningless to say that the divergences are of the form $(1/\e^n) {\bar D}^2 (\G^{\hat{\hat{\m}}}\G_{\hat{\hat{\m}}}) $
instead of $(1/\e^n) W^\a W_\a$, since we have the relation (\ref{eq:wwtotder}). In fact, this statement is quite  nontrivial.
The divergent pole $(1/\e^n){\bar D}^2 (\G^{\hat{\hat{\m}}}\G_{\hat{\hat{\m}}})  $ arises from $n$ or higher loop diagrams, 
and hence the coefficient $c_n(g)$ is of the form $c_n(g) \sim g^{2n}+\cdots$.
This means that the divergent pole $(1/\e^n) W^\a W_\a$ comes from $n+1$ or higher loops and has a coefficient of order $g^{2(n+1)}$.
This is suppressed by additional $g^2$ than naively expected. In fact, this suppression can be confirmed by the two-loop calculation of 
Grisaru, Milewski and Zanon~\cite{Grisaru:1985yk,Grisaru:1985ik}. In our notation and convention, their result is summarized as~\footnote{
One should note that what they called the bare operator $W^{(0)}_\a$ is in fact $(g/g_0) W_\a$ in our notation, 
where $g_0$ is the bare gauge coupling in DRED. The operator $W_\a$ in their notation coincides with our $W_\a$
because of the relation $Z_V=(g/g_0)^2$, where $Z_V$ is defined in their paper. The supercurrent should have been defined as
$(1/g^2_0) W_\a \bar{W}_{\dot{\a}}+\cdots$ instead of $W_\a^{(0)} \bar{W}^{(0)}_{\dot \a}+\cdots$.  }
\beq
[W^\a W_\a]_{\rm DR}=\left(1-\frac{3}{\e} \left(\frac{t_{Ad}g^2}{16\pi^2}\right)^2 +{\cal O}(g^6)\right)W^\a W_\a.
\eeq
Having confidence that the divergence is indeed $(1/\e^n){\bar D}^2 (\G^{\hat{\hat{\m}}}\G_{\hat{\hat{\m}}}) $,
we suspect that there exists a ``hidden divergence'' of the form $(1/\e){\bar D}^2 (\G^{\hat{\hat{\m}}}\G_{\hat{\hat{\m}}})$, which
should be subtracted by nonzero $c_1(g)$. (In fact, Eq.~(\ref{eq:epsilonscalarcounterterm}) is a similar kind of such hidden divergence which becomes
clear by the analytic continuation of parameters into superspace~\cite{ArkaniHamed:1998kj}.)
This subtraction will give a different normalization of the renormalized operator $[W^\a W_\a]$
than the operator $[W^\a W_\a]_{\rm DR}$ subtracted by DR.

The ${\cal N}=4$ regularization avoids the above difficulty by effectively replacing the counterterms as
\beq
\sum_{n=1}^\infty \frac{c_n(g)}{\e^n} \frac{1}{4}\bar{D}^2 (\G^{\hat{\hat{\m}}}\G_{\hat{\hat{\m}}}) \to {\bar D}^2 K,
\eeq
where $K$ is defined in Eq.~(\ref{eq:wwcounterterm}).
In this way, the structure of the divergence becomes much more clear. We can uniquely specify the overall normalization of $[W^\a W_\a]$, 
which is preferred by the topological nature of $F_{\m\n}\tilde{F}^{\m\n}$. 
(Strictly speaking, we have regularized the ${\cal N}=4$ theory by DRED to make our arguments concrete. 
But we believe that this is not essential. Any other regularization is 
possible as long as the arguments of section~\ref{sec:4} make sense.)
Our renormalization prescription for $W^\a W_\a$ gives the one-loop exact coefficient in the trace anomaly. 

As to the second supercurrent ${\cal J}^{(2)}_{\a\dot{\a}}$, our guess is that this supercurrent is simply related to the first one
by
\beq
{\cal J}^{(2)}_{\a\dot{\a}}=\frac{\b_{\rm one-loop}}{\b_{\rm two-loop}}{\cal J}^{(1)}_{\a\dot{\a}}+[{\rm evanescent}], \label{eq:twoone}
\eeq
where by the term [evanescent], we mean an operator which vanishes in the limit $\e \to 0$ when it is inserted into a renormalized correlation function.
In dimensional regularization or reduction, it often happens that two different-looking operators are in fact the same 
up to evanescent terms (see e.g., Ref.~\cite{Collins:1984xc}).
In fact, in the construction of the two supercurrents, operators such as $[D_\a ,\bar{D}_\aa][\G^{\hat{\hat{\m}}}\G_{\hat{\hat{\m}}}]_{\rm DR}$
are necessary which seem to be evanescent operators.
We do not perform a detailed study on this point, but the existence of the two different supercurrents cannot be confirmed 
unless the possibility (\ref{eq:twoone}) is excluded.
(However, the idea of two supercurrents becomes important when matter fields are included
in the theory~\cite{Yonekura:2010mc,Huang:2010tn,Ensign:1987wy}.)

Next we discuss the result obtained by Arkani-Hamed and Murayama~\cite{ArkaniHamed:1997mj}.
They computed the anomalies in dilatation transformations. Their result can be summarized as follows.
If we take the dilatation transformation of the gauge superfield $V(x,\h,\bar{\h})$ as
\beq
V(x,\h,\bar{\h}) \to V(e^{t} x, e^{t/2}\h,e^{t/2}\bar{\h}),
\eeq
where $t$ is the parameter of the transformation, we obtain the anomaly given by
\beq
\int d^4 x \int d^2\h \left( t \frac{3t_{Ad}}{32\pi^2}  \right) W^\a W_\a+{\rm h.c.}. \label{eq:firstdilatation}
\eeq
On the other hand, we can canonically normalize the gauge field as $V=g_c V_c$, where $g_c$ is the ``canonical gauge coupling'',
and  perform the dilatation transformation as
\beq
g_c V_c (x,\h,\bar{\h}) \to g'_c(t)V_c(e^{t} x, e^{t/2}\h,e^{t/2}\bar{\h}), \label{eq:canonicaldilatation}
\eeq
where $g'_c(t)$ is chosen so that the gauge field is canonically normalized after the dilatation transformation.
Then the anomaly is given by
\beq
\int d^4 x \int d^2\h \frac{1}{4}\left( t \frac{3t_{Ad}}{8\pi^2}-\frac{t_{Ad}}{8\pi^2}\log \left(\frac{g'^2_c(t)}{g^2_c}\right) \right) W^\a(g'_c(t)V_c) W_\a(g'_c(t)V_c) +{\rm h.c.}.
\eeq
Here $g'_c(t)$ is determined by the equation
\beq
\frac{1}{g'^2_c(t)}=\frac{1}{g_c^2}+ t \frac{3t_{Ad}}{8\pi^2}-\frac{t_{Ad}}{8\pi^2}\log \left(\frac{g'^2_c(t)}{g^2_c}\right).
\eeq
Taking $t$ infinitesimal, we get
\beq
\frac{1}{g'^2_c(t)}=\frac{1}{g_c^2}-t \frac{\b_{\rm NSVZ}}{g_c^4}
\eeq
where $\b_{\rm NSVZ}$ is the NSVZ beta function.
In the transformation (\ref{eq:canonicaldilatation}), the classical action also changes.
The total change of the path integral $\int [D V]e^{iS}$ in this case is given by
\beq
\int d^4 x \int d^2\h t\b_{\rm NSVZ} \left. \frac{\q}{\q g_c^2}  \right|_{V_c} \left( \frac{1}{4g^2_c} W^\a(g_c V_c) W_\a(g_cV_c) \right)
+{\rm h.c.}. \label{eq:seconddilatation}
\eeq
Therefore, it was claimed that there are two dilatation anomalies. One is given by Eq.~(\ref{eq:firstdilatation}), which has the one-loop exact form,
while the other is given by Eq.~(\ref{eq:seconddilatation}) which, at first slight, has higher order corrections.
This result may seem to contradict with our result, since we have obtained the unique trace anomaly, up to a total derivative term
and the ambiguity of choosing $C_I$ discussed in~\ref{app:A}.

However, a closer look at Eq.~(\ref{eq:seconddilatation}) gives us the resolution to this problem. We can rewrite this equation as
\beq
t \b_{\rm NSVZ}\left. \frac{\q S}{\q g_c^2} \right|_{V_c}= t\b_{\rm NSVZ} \left(\left. \frac{\q S}{\q g_c^2} \right|_{V}
 + \int d^4 x d^2\h d^2 \bar{\h}  \frac{1}{2g_c^2}\frac{\d S}{\d V}V \right), \label{eq:partialder}
\eeq
where $S$ is the action and $\d S/\d V$ is the functional differentiation of the action with respect to $V$.
Naively, $V(\d S/\d V)$ vanishes by equations of motion. However, the computation of the anomaly 
under the rescaling $V \to ({\rm const.}) V$ performed by Arkani-Hamed and Murayama indicates that the equation of motion in their regularization scheme
should really be given as (see \ref{app:A}, especially Eqs.~(\ref{eq:rescalingfield}),~(\ref{eq:numbercount}) and (\ref{eq:countingoperator})),
\beq
0 \approx N_V(x,\h, \bar{\h}) \equiv  \frac{\d S}{\d V}V-\frac{t_{Ad}}{16 \pi^2} \left(-\frac{1}{2}(e^{2V}D^\a e^{-2V}) W_\a \right) +{\rm h.c.}  +({\rm total~derivative}) ,
\eeq
where one should note that the second term satisfies
\beq
-\frac{1}{4} \bar{D}^2\left(-\frac{1}{2}(e^{2V}D^\a e^{-2V}) W_\a \right)=W^\a W_\a.
\eeq
Then, Eq.~(\ref{eq:partialder}) becomes
\beq
&&t\b_{\rm NSVZ} \int d^4 x d^2 \h \left(-\frac{1}{4g_c^4}+\frac{t_{Ad}}{32 \pi^2 g_c^2} \right) W^\a W_\a+{\rm h.c.}
+\int d^4 x d^2\h d^2\bar{\h} t\frac{\b_{\rm NSVZ}}{2g_c^2}N_V \nonumber \\
&=&\int d^4 x \int d^2\h \left( t\frac{3t_{Ad}}{32\pi^2}  \right) W^\a W_\a+{\rm h.c.}+\int d^4 x d^2\h d^2\bar{\h} t\frac{\b_{\rm NSVZ}}{2g_c^2}N_V.
\eeq
Therefore, the difference of the two anomalies is only the term containing $N_V$, which vanishes by the equation of motion.
This just reflects the ambiguity of the definition of the energy-momentum tensor discussed in \ref{app:A}.
In fact, by setting $\d x^\m=x^\m$ in Eqs.~(\ref{eq:matterneothertransf}) and (\ref{eq:emwtid}) of \ref{app:A}, we can see that
the two dilatation anomalies computed above are just the trace $T^\m_\m$ with different values of $C_I$.
In this way, we obtain the one-loop exact coefficient for the operator $W^\a W_\a$ in the dilatation anomaly, which is consistent with our result.

Shifman and Vainshtein~\cite{Shifman:1986zi} gave a supercurrent equation which is consistent with our result on the trace anomaly.
Although there is no contradiction in the supercurrent equation, their interpretation of the anomaly puzzle is different from ours.
They argued that the one-loop contribution to the beta function comes from UV region, but higher loop effects are from IR region.
Since operator equations are given in terms of UV quantities, they claimed that the one-loop beta function should appear in the trace anomaly.
However, as emphasized by Arkani-Hamed and Murayama~\cite{ArkaniHamed:1997mj}, 
both the one-loop beta function and the NSVZ beta function can be derived
by only using the information of UV physics. 
Indeed, the derivation of the NSVZ beta function we reviewed in section~\ref{sec:4} has used only the decoupling 
of the adjoint chiral fields, Eq.~(\ref{eq:lowenergycoupling}), which is completely determined in UV region.\footnote{We have not even used the 
Jacobian from the path-integral measure as in Ref.~\cite{ArkaniHamed:1997mj}. }
Then the question 
of which beta function should appear in the trace anomaly comes back to us.
This is precisely the problem we have studied in this paper. Although there are many renormalization schemes which give different beta functions, 
only one scheme possesses the desired property
\beq
\frac{\q}{\q g^{-2}} \vev{ (\cdots)}=i\vev{\left(\int d^4x d^2\h\frac{1}{4}[W^\a W_\a]_R+{\rm h.c.}\right)(\cdots)}
\eeq
for the appropriately renormalized operator $[W^\a W_\a]_R$. Without this property, the trace anomaly formula is simply false.

\section*{Acknowledgements}
The author is grateful to Y.~Nakayama and especially T.~Moroi for stimulating discussions. 
This work was supported by 
World Premier International Research Center Initiative
(WPI Initiative), MEXT, Japan,
and supported
in part by JSPS Research Fellowships for Young Scientists.

\appendix
\setcounter{equation}{0}
\renewcommand{\theequation}{\Alph{section}.\arabic{equation}}
\renewcommand{\thesection}{Appendix~\Alph{section}}

\section{Energy-momentum tensor}\label{app:A}
\setcounter{equation}{0}
In this appendix we review basic properties of the energy-momentum tensor.
The energy-momentum tensor can be defined as the Noether current of the Poincare symmetry.
Let us consider a theory described by fields $\v_I(x)$ which are in some representations of the Lorentz group.
We treat $\v_I$ as if they are bosons for simplicity, but a generalization to fermions is obvious.
An infinitesimal transformation of the coordinates is given as 
\beq
x^\mu \to x'^\m= x^\mu+\d x^\mu(x).
\eeq
We define a transformation of the field $\v_I$ under the above coordinate transformation as
\beq
\v_I(x) &\to& \v'_I(x) = \v_I(x)+\d\v_I(x), \\
\d\v_I(x)&=& \d x^\mu \q_\mu \v_I(x)+\frac{1}{2}\q_{[\m}\d x_{\n]} \S^{\m\n}_I\v_I(x) + \q_\m \d x^\m C_I \v_I(x), \label{eq:matterneothertransf}
\eeq 
where $\S^{\mu\nu}_I$ is the Lorentz symmetry generator acting on $\v_I$, and $C_I$ is an arbitrary constant.
This definition is taken so that it coincides with Poincare transformation when $\d x^\m$ is given as $\d x^\mu=\w^\m_{~\n} x^\n+a^\m$,
where $a_\m$ and $\w_{\m\n}=-\w_{\n\m}$ are constants.

Let us consider a correlation function
\beq
\vev{\prod_i\v_{I_i}(x_i)} =\int [D\v] e^{iS[\v]} \prod_i\v_{I_i}(x_i) . \label{eq:correlation}
\eeq
Then, by performing the above transformation of the fields in the path integral, we obtain
\beq
0=-\vev{ i \int d^d x  T^{\m\n} \q_\m \d x_\n  \prod_i\v_{I_i}(x_i) } +\sum_i \vev{ \d \v_{I_i} (x_i)\prod_{j \neq i} \v_{I_j}(x_j)}, \label{eq:emwtid}
\eeq
where $d$ is the space-time dimension.
This is the defining equation of the energy-momentum tensor,
but a slight ambiguity remains.
There is a freedom to change $T^{\m\n}$ as $T^{\m\n}+\q_\r A^{\r\m\n}$, where $A^{\r\m\n}$ is an arbitrary tensor satisfying
$A^{\r\m\n}=-A^{\m\r\n}$. We can use this freedom to make $T^{\m\n}$ symmetric, $T^{\m\n}=T^{\n\m}$. 
First, note that if $\d x^\m=\w^\mu_{~\n} x^\n$ where 
$\w_{\m\n}$ is an antisymmetric constant tensor, the path integral measure times the factor $e^{iS[\v]}$
is invariant by the Lorentz symmetry. Then $\int d^dx T^{[\m\n]} \w_{\m\n}=0 $ for an arbitrary constant 
$\w_{\m\n}$, and hence $T^{[\m\n]}$ is a total derivative, $T^{[\m\n]}=\q_\r B^{\r\m\n}$. We can define a symmetric energy-momentum tensor as
$T^{\m\n}-\q_\r (B^{\r\m\n}-B^{\m\r\n}-B^{\n\r\m})$, i.e., $A^{\r\m\n}=-B^{\r\m\n}+B^{\m\r\n}+B^{\n\r\m}$. In the following, we use this symmetric tensor.
Even after making $T^{\m\n}$ symmetric, there still remains a freedom to change $T^{\m\n}$ as
\beq
T^{\m\n} \to T^{\m\n}+\q_\r \q_\s C^{\m\r\n\s},
\eeq
where $C_{\m\r\n\s}$ is an arbitrary field satisfying $C_{\m\r\n\s}=-C_{\r\m\n\s}=C_{\n\s\m\r}$. This is the so-called improvement of the energy-momentum
tensor. If one prefers to define $T^{\m\n}$ by functional differentiation of the action with respect to the metric tensor $g_{\m\n}$,
this freedom corresponds to adding a term $-\frac{1}{4}\int d^d R_{\m\r\n\s}C^{\m\r\n\s}$ to the action, where $R_{\m\r\n\s}$ is the Riemann tensor.

Functionally differentiating Eq.~(\ref{eq:emwtid}) by $\d x^\n(y)$, we get the Ward-Takahashi identity
\beq
&& \vev{\q_\m T^{\m\n} (y) \prod_i\v_{I_i}(x_i) } = \nonumber \\
&& i\sum_i 
\vev{ \left(\d^{(d)}(x_i-y)\q^\n \v_{I_i} (x_i)-\left(\frac{1}{2} \S^{\m\n}_{I_i}+C_{I_i}\y^{\mu\n}\right) (\q_y)_\m\d^{(d)}(x_i-y) \v_{I_i}(x_i) 
\right)\prod_{j \neq i} \v_{I_j}(x_j)}. \nonumber \\ \label{eq:WTidentityofEMtensor} 
\eeq
In particular, if $y \neq x_i$, this equation is just the conservation of the energy-momentum tensor, $\q_\m T^{\m\n} \approx 0$, where $\approx$ means
that the equation is valid up to contact terms.

Aside from the improvement, there exists the ambiguity of choosing $C_I$ in the definition of the energy-momentum tensor.
To investigate this point, let us consider a rescaling transformation of the field $\v_I$ as 
\beq
\v_I(x)  \to (1+\a(x))\v_I(x), \label{eq:rescalingfield}
\eeq
where $\a(x)$ is an infinitesimal function. By performing this rescaling in the correlation function~(\ref{eq:correlation}) and functionally differentiating 
the result by $\a(y)$, we obtain
\beq
0=\vev{ i N_I(y) \prod_i\v_{I_i}(x_i) } +\sum_{I_i=I} \vev{ \d^{(d)}(x_i-y) \v_{I_i}(x_i)\prod_{j \neq i} \v_{I_j}(x_j)}. \label{eq:numbercount}
\eeq
Here we have defined the operator $N_I$ as
\beq
N_I(x)= \frac{\d S[\v]}{\d \v_I(x)}\v_I(x) +A_I(x), \label{eq:countingoperator}
\eeq
where $A_I(x)$ is the anomaly arising from the path integral measure in the transformation~(\ref{eq:rescalingfield}).
The operator $N_I$ vanishes by equations of motion, $N_I \approx 0$, since there are only contact terms in Eq.~(\ref{eq:numbercount}) aside from 
the term containing $N_I$.
Using $N_I$, one can see that the ambiguity in the definition of the energy-momentum tensor represented by $C_I$ is 
of the form
\beq
T^{\m\n}=T^{\m\n}|_{C_I=0}-\y^{\m\n}\sum_I C_IN_I.
\eeq
Thus, energy-momentum tensors with different values of $C_I$ are the same up to the equations of motion, $N_I \approx 0$.
More generally, any symmetric tensor which vanishes by equations of motion can be added to $T^{\m\n}$ without violating the conservation
equation $\q_\m T^{\m\n} \approx 0$. One can also check that this ambiguity does not affect the generators of 
the Poincare symmetry $P^\m=\int d^{d-1}x T^{0\m}$ and $J^{\m\n}=\int d^{d-1}x (x^\m T^{0\n}-x^\n T^{0\m})$ 
when Eq.~(\ref{eq:WTidentityofEMtensor}) is integrated to give 
the commutation relations $[P_\m, \v_I]=i\q_\m \v_I$ and $[J_{\m\n},\v_I]=i(x_\m \q_\n-x_\n \q_\m + \S_{I\m\n})\v_I$.

Finally, let us comment on the role of $C_I$ at the fixed point of RG flow. 
In section~\ref{sec:2}, we derive the formula for the trace of the energy-momentum tensor given by Eq.~(\ref{eq:traceanomalyformula}).
By using that formula, the dilatation current
\beq
j_D^\m =x_\n T^{\m\n}-j_V^\m
\eeq
can be shown to satisfy the Ward-Takahashi identity at the fixed point $d_a \l_a-\beta_a=0$,
\beq
&&\vev{ \q_\m j^\m_D (y) \prod_i\v_{I_i}(x_i)}= \nonumber \\
&&i\sum_i 
\vev{\d^{(d)}(x_i-y) \left(x_i^\n\q_\n \v_{I_i} (x_i)+\D_{I_i} \v_{I_i}(x_i) 
\right)\prod_{j \neq i} \v_{I_j}(x_j)} \nonumber \\
&&+i(\q_y)_\m \sum_i 
 \vev{\d^{(d)}(x_i-y) \left(-\frac{1}{2} y_\n \S^{\m\n}_{I_i}- y^\m C_{I_i}   
\right)\v_{I_i}(x_i)\prod_{j \neq i} \v_{I_j}(x_j)}. \label{eq:dilatationWTid}
\eeq
If we integrate this equation over $y^\m$ to obtain the integrated version of the Ward-Takahashi identity, the final line of Eq.~(\ref{eq:dilatationWTid})
does not contribute at all. Thus the arbitrary parameter $C_I$ has no physical meaning in the scaling symmetry.


\begin{thebibliography}{99}


\bibitem{Ferrara:1974pz}
  S.~Ferrara and B.~Zumino,
  Nucl.\ Phys.\  B {\bf 87}, 207 (1975).
  
\bibitem{Clark:1978jx}
  T.~E.~Clark, O.~Piguet and K.~Sibold,
  Nucl.\ Phys.\  B {\bf 143}, 445 (1978).
  
\bibitem{Grisaru:1978vx} 
  M.~T.~Grisaru,
  In *Shifman, M.A. (ed.): The many faces of the superworld* 370-387
  
\bibitem{Adler:1969er}
  S.~L.~Adler and W.~A.~Bardeen,
  Phys.\ Rev.\  {\bf 182}, 1517 (1969).
  
\bibitem{Grisaru:1985yk} 
  M.~T.~Grisaru, B.~Milewski and D.~Zanon,
  Phys.\ Lett.\ B {\bf 157}, 174 (1985).
  
\bibitem{Grisaru:1985ik} 
  M.~T.~Grisaru, B.~Milewski and D.~Zanon,
  Nucl.\ Phys.\ B {\bf 266}, 589 (1986).
  
   \bibitem{Shifman:1986zi}
  M.~A.~Shifman and A.~I.~Vainshtein,
  Nucl.\ Phys.\  B {\bf 277}, 456 (1986)
  [Sov.\ Phys.\ JETP {\bf 64}, 428 (1986\ ZETFA,91,723-744.1986)];

  \bibitem{ArkaniHamed:1997mj}
  N.~Arkani-Hamed and H.~Murayama,
  JHEP {\bf 0006}, 030 (2000)
  [arXiv:hep-th/9707133].
  

  
  
\bibitem{Komargodski:2010rb}
  Z.~Komargodski and N.~Seiberg,
  JHEP {\bf 1007}, 017 (2010)
  [arXiv:1002.2228 [hep-th]].
  
\bibitem{Magro:2001aj}
  M.~Magro, I.~Sachs and S.~Wolf,
  Annals Phys.\  {\bf 298}, 123 (2002)
  [arXiv:hep-th/0110131].

\bibitem{Yonekura:2010mc} 
  K.~Yonekura,
  JHEP {\bf 1009}, 049 (2010)
  [arXiv:1004.1296 [hep-th]].

\bibitem{Huang:2010tn} 
  X.~Huang and L.~Parker,
  Eur.\ Phys.\ J.\ C {\bf 71}, 1570 (2011)
  [arXiv:1001.2364 [hep-th]].

  
\bibitem{Novikov:1983uc}
  V.~A.~Novikov, M.~A.~Shifman, A.~I.~Vainshtein and V.~I.~Zakharov,
  Nucl.\ Phys.\  B {\bf 229}, 381 (1983);
  
\bibitem{Jack:1996cn} 
  I.~Jack, D.~R.~T.~Jones and C.~G.~North,
  Nucl.\ Phys.\ B {\bf 486}, 479 (1997)
  [hep-ph/9609325].

\bibitem{Adler:1976zt} 
  S.~L.~Adler, J.~C.~Collins and A.~Duncan,
  Phys.\ Rev.\ D {\bf 15}, 1712 (1977).
  
\bibitem{Collins:1976yq} 
  J.~C.~Collins, A.~Duncan and S.~D.~Joglekar,
  Phys.\ Rev.\ D {\bf 16}, 438 (1977).
  
\bibitem{Nielsen:1977sy} 
  N.~K.~Nielsen,
  Nucl.\ Phys.\ B {\bf 120}, 212 (1977).
  
\bibitem{Dine:2011gd} 
  M.~Dine, G.~Festuccia, L.~Pack, C.~-S.~Park, L.~Ubaldi and W.~Wu,
  JHEP {\bf 1105}, 061 (2011)
  [arXiv:1104.0461 [hep-th]].



\bibitem{Siegel:1979wq}
  W.~Siegel,
  Phys.\ Lett.\  B {\bf 84}, 193 (1979).
  
\bibitem{Capper:1979ns}
  D.~M.~Capper, D.~R.~T.~Jones and P.~van Nieuwenhuizen,
  Nucl.\ Phys.\  B {\bf 167}, 479 (1980).
  
\bibitem{Stockinger:2005gx} 
  D.~Stockinger,
  JHEP {\bf 0503}, 076 (2005)
  [hep-ph/0503129].

  
\bibitem{Lowenstein:1971jk} 
  J.~H.~Lowenstein,
  Commun.\ Math.\ Phys.\  {\bf 24}, 1 (1971).
  

\bibitem{Breitenlohner:1977hr} 
  P.~Breitenlohner and D.~Maison,
  Commun.\ Math.\ Phys.\  {\bf 52}, 11 (1977).
  
 
  
  
\bibitem{Wess:1992cp}
  J.~Wess and J.~Bagger,
{\it  Princeton, USA: Univ. Pr. (1992) 259 p}

\bibitem{Seiberg:1993vc} 
  N.~Seiberg,
  Phys.\ Lett.\ B {\bf 318}, 469 (1993)
  [hep-ph/9309335].
  
\bibitem{ArkaniHamed:1998kj} 
  N.~Arkani-Hamed, G.~F.~Giudice, M.~A.~Luty and R.~Rattazzi,
  Phys.\ Rev.\ D {\bf 58}, 115005 (1998)
  [hep-ph/9803290].
  
\bibitem{Grisaru:1985tc}
  M.~T.~Grisaru, B.~Milewski and D.~Zanon,
  Phys.\ Lett.\  B {\bf 155}, 357 (1985).
  
\bibitem{Argyres:1996eh} 
  P.~C.~Argyres, M.~R.~Plesser and N.~Seiberg,
  Nucl.\ Phys.\ B {\bf 471}, 159 (1996)
  [hep-th/9603042].

  
\bibitem{Callan:1970ze} 
  C.~G.~Callan, Jr., S.~R.~Coleman and R.~Jackiw,
  Annals Phys.\  {\bf 59}, 42 (1970).
  
\bibitem{Espriu:1982bw} 
  D.~Espriu and R.~Tarrach,
  Z.\ Phys.\ C {\bf 16}, 77 (1982).
  
\bibitem{Breitenlohner:1983pi} 
  P.~Breitenlohner, D.~Maison and K.~S.~Stelle,
  Phys.\ Lett.\ B {\bf 134}, 63 (1984).
  
\bibitem{Collins:1984xc} 
  J.~C.~Collins,
  ``Renormalization. An Introduction To Renormalization, The Renormalization Group, And The Operator Product Expansion,''
  Cambridge, Uk: Univ. Pr. ( 1984) 380p

  
 
  
\bibitem{Ensign:1987wy} 
  P.~Ensign and K.~T.~Mahanthappa,
  Phys.\ Rev.\ D {\bf 36}, 3148 (1987).




\end{thebibliography}
\end{document}